\documentclass[usenatbib]{mn2e}

\date{Accepted 5 August 2009}

\usepackage{txfonts}
\usepackage{graphicx}

\newcommand{\be}{\begin{equation}}
\newcommand{\ee}{\end{equation}}
\newcommand{\bea}{\begin{eqnarray}}
\newcommand{\eea}{\end{eqnarray}}
\newcommand{\ha}{HI}
\newcommand{\hm}{H$_2$}
\newcommand{\avg}[1]{\langle#1\rangle}

\newcommand{\xco}{X_{\rm CO}}

\newcommand{\f}{R_{\rm mol}}
\newcommand{\fc}{\f^{\rm c}}

\newcommand{\rd}{r_{\rm disc}}
\newcommand{\rha}{r_{\rm HI}}
\newcommand{\rhm}{r_{\rm H_2}}
\newcommand{\rhahalfmass}{\rha^{\rm half}}
\newcommand{\rhmhalfmass}{\rhm^{\rm half}}

\newcommand{\rvir}{r_{\rm vir}}
\newcommand{\mass}{M}
\newcommand{\mg}{\mass_{\rm gas}}
\newcommand{\mha}{\mass_{\rm HI}}
\newcommand{\mhm}{\mass_{{\rm H}_2}}

\newcommand{\ms}{\mass_{\rm s}}
\newcommand{\msdisk}{\mass_{\rm s}^{\rm disc}}
\newcommand{\mvir}{\mass_{\rm vir}}

\newcommand{\Sigmaha}{\Sigma_{\rm HI}}
\newcommand{\Sigmahm}{\Sigma_{\rm H_2}}
\newcommand{\Sigmahamax}{\Sigmaha^{\rm max}}
\newcommand{\Sigmahmmax}{\Sigmahm^{\rm max}}
\newcommand{\Sigmahafwhm}{\Sigmaha^{\rm FWHM}}
\newcommand{\Sigmahmfwhm}{\Sigmahm^{\rm FWHM}}
\newcommand{\Sigmahahalfmass}{\Sigmaha^{\rm half}}
\newcommand{\Sigmahmhalfmass}{\Sigmahm^{\rm half}}
\newcommand{\Sigmastd}{\tilde{\Sigma}_{\rm H}}
\newcommand{\Omegaha}{\Omega_{\rm HI}}
\newcommand{\Omegahm}{\Omega_{{\rm H}_2}}
\newcommand{\msun}{{\rm M}_{\odot}}
\newcommand{\msunpc}{\msun{\rm\,pc^{-2}}}
\newcommand{\kpc}{{\rm kpc}}

\newcommand{\mnras}{MNRAS}
\newcommand{\nat}{Nat}
\newcommand{\aj}{AJ}
\newcommand{\apj}{ApJ}
\newcommand{\apjl}{ApJL}
\newcommand{\aap}{A\&A}

\title{Compactness of Cold Gas in High-Redshift Galaxies}

\author[D. Obreschkow and S. Rawlings]{D. Obreschkow and S. Rawlings\\
Astrophysics, Department of Physics, University of Oxford, Keble Road, Oxford, OX1 3RH, UK}

\begin{document}

\maketitle

\label{firstpage}

\begin{abstract}
Galaxies in the early Universe were more compact and contained more molecular gas than today. In this paper, we revisit the relation between these empirical findings, and we quantitatively predict the cosmic evolution of the surface densities of atomic (\ha) and molecular (\hm) hydrogen in regular galaxies. Our method uses a pressure-based model for the \hm/\ha-ratio of the Interstellar Medium, applied to $\sim\!3\cdot10^7$ virtual galaxies in the Millennium Simulation. We predict that, on average, the \ha-surface density of these galaxies saturates at $\Sigmaha<10~\msunpc$ at all redshifts ($z$), while \hm-surface densities evolve dramatically as $\Sigmahm\propto(1+z)^{2.4}$. This scaling is dominated by a $\propto(1+z)^2$ surface brightness scaling originating from the $\propto(1+z)^{-1}$ size scaling of galaxies at high $z$. Current measurements of $\Sigmahm$ at high $z$, derived from CO-observations, tend to have even higher values, which can be quantitatively explained by a selection bias towards merging systems. However, despite the consistency between our high-$z$ predictions and the sparse empirical data, we emphasize that the empirical data potentially suffer from serious selection biases and that the semi-analytic models remain in many regards uncertain. As a case study, we investigate the cosmic evolution of simulated galaxies, which resemble the Milky Way at $z=0$. We explicitly predict their \ha- and \hm-distribution at $z=1.5$, corresponding to the CO-detected galaxy $BzK$-21000, and at $z=3$, corresponding to the primary science goal of the Atacama Large Millimeter/submillimeter Array (ALMA).
\end{abstract}


\section{Introduction}\label{section_introduction}

Galaxies were more compact in the early Universe than today \citep{Bouwens2004,Trujillo2006,Buitrago2008}. This empirical feature is probably driven by an increase in the volume-to-mass ratio of dark haloes with cosmic time \citep{Gunn1972}, which dictates the density evolution of galaxies by the transfer of angular momentum \citep{Fall1980}. We have recently argued \citep{Obreschkow2009c} that the density evolution of galactic discs implies a systematic pressure change, which causes a dramatic decline in the mass ratio between molecular (\hm) and atomic (\ha) hydrogen with cosmic time. Using the Millennium Simulation, we showed that this decline in the \hm/\ha-ratio simultaneously explains (i) the observations of large molecular masses in ordinary galaxies at $z=1.5$ \citep{Daddi2008}, (ii) the weak cosmic evolution of the \ha-density $\Omegaha$ inferred from damped Lyman-$\alpha$ systems \citep{Prochaska2005}, and (iii) the history of star formation inferred from ultraviolet, far-infrared, submillimeter, and radio continuum observations \citep{Hopkins2006}.

In this paper, we quantitatively predict the cosmic evolution of the surface densities of \ha~and \hm~in a large sample of regular galaxies. Section \ref{section_model} explains our physical model and the numerical simulation. Section \ref{section_global_evolution} presents our predictions for the average cosmic evolution of the \ha- and \hm-distributions in a broad sample of galaxies, while Section \ref{section_mw_evolution} specifically focusses on the evolution of \ha- and \hm-distributions in Milky Way (MW)-type galaxies. Section \ref{section_conclusion} gives a brief summary and outlook.

\section{Simulation of $\Sigmaha$ and $\Sigmahm$ in galaxies}\label{section_model}

\begin{figure}
\includegraphics[width=\columnwidth]{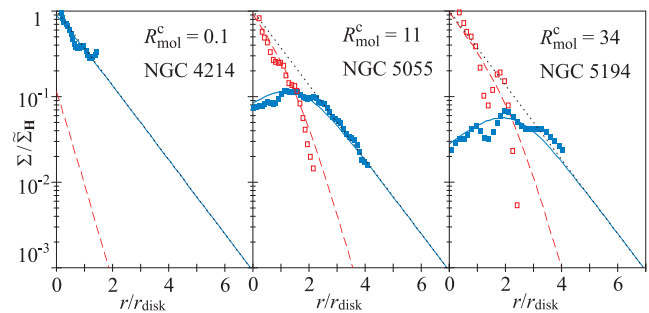}
\caption{\small{Surface densities of galactic cold gas as a function of galactic radius. Lines show the predictions of Eqs.~(\ref{eqsigmaha}, \ref{eqsigmahm}) for $\Sigmaha(r)$ (solid), $\Sigmahm(r)$ (dashed), and $\Sigmaha(r)+\Sigmahm(r)$ (dotted) for different values of $\fc$. Dots represent the observed counterparts \citep{Leroy2008} for $\Sigmaha(r)$ (filled) and $\Sigmahm(r)$ (empty), inferred from CO(2--1) observations.}}
\label{fig_fc_solutions}
\end{figure}

Several computational studies \citep[e.g.][]{Combes1999,Greve2008,Robertson2008,Dobbs2008} have investigated the distribution of molecular gas in regular galaxies and revealed pivotal connections between the distribution of molecules and star formation. In particular, \citet{Robertson2008} and \citealp{Krumholz2009} (see also \citealp{Elmegreen1993}) demonstrated that the empirical relation between the local \hm/\ha-ratios and cold gas pressure, measured by \citet{Blitz2006} and \citet{Leroy2008}, can be approximately reproduced by SPH/$N$-body simulations, which include a model for interstellar radiation.

While those simulations are effective tools for the study of the cold gas in individual galaxies, they are computationally too expensive to be included in ``cosmological simulations'', i.e.~simulations of representative samples of galaxies with resolved merger histories in an expanding model-universe. However, cosmological simulations of \ha- and \hm-distributions are required for the design and analysis of high-$z$ cold gas surveys possible with future telescopes, such as the Square Kilometer Array (SKA), the Large Millimeter Telescope (LMT), and the Atacama Large Millimeter/submillimeter Array (ALMA). To circumvent the current computational bottleneck, we decided \citep{Obreschkow2009b} to adopt a semi-analytic simulation of millions of galaxies and to estimate the \ha-~and \hm-distributions of these galaxies using an analytic model. We shall first explain this model and then the semi-analytic simulation.

Our model for the distributions of \ha~and \hm~in regular galaxies represents a ramification of the \hm/\ha--pressure relation measured by \citet{Leroy2008}, combined with two assumptions for the distribution of cold hydrogen. The first assumption is that all galaxies carry their cold gas in flat discs, such as is observed in virtually all regular galaxies (see \citet{Leroy2008} for spiral galaxies at low $z$, \citet{Young2002} for elliptical galaxies at low $z$, and e.g.~\citet{Tacconi2006} for galaxies at high $z$). The second assumption is that the surface density of cold hydrogen, $\Sigmaha(r)+\Sigmahm(r)$, follows an exponential profile in the radial coordinate $r$ \citep[e.g.][]{Leroy2008}. Based on these assumptions we have shown \citep{Obreschkow2009b} that the surface densities of \ha~and \hm~are
\bea
  \Sigmaha(r) & = & \frac{\Sigmastd\,\exp(-r/\rd)}{1+\fc\exp(-1.6\,r/\rd)}\ , \label{eqsigmaha} \\
  \Sigmahm(r) & = & \frac{\Sigmastd\,\fc\,\exp(-2.6\,r/\rd)}{1+\fc\exp(-1.6\,r/\rd)}\ , \label{eqsigmahm}
\eea
where $\rd$ is the exponential scale radius of the cold gas disc, $\Sigmastd=(\mha+\mhm)/(2\pi\rd^2)$ is a normalization factor, and $\fc=\Sigmahm(0)/\Sigmaha(0)$ is the central \hm/\ha-ratio. $\fc$ can be approximated \citep{Obreschkow2009b} from $\rd$, the disc stellar mass $\msdisk$, and the total (\ha+\hm+He) cold gas mass $\mg$ as,
\be\label{eqfc}
  \fc =\!\left[K\,\rd^{-4}\mg\big(\mg\!+\!0.4\,\msdisk\big)\right]^{0.8},
\ee
where $K=11.3\rm\,m^4\,kg^{-2}$ is an empirical constant.

If the radii and surface densities in Eqs.~(\ref{eqsigmaha}, \ref{eqsigmahm}) are normalized to $\rd$ and $\Sigmastd$, then $\Sigmaha(r)$ and $\Sigmahm(r)$ become unique functions of $\fc$. Fig.~\ref{fig_fc_solutions} shows $\Sigmaha(r)$ and $\Sigmahm(r)$ in these normalized coordinates for three different values of $\fc$, chosen to match those of nearby galaxies with measured \ha- and CO-densities \citep{Leroy2008}: NGC 4214, a star-forming, \ha-rich, dwarf galaxy; NGC 5055, a massive spiral galaxy with similar \ha- and \hm-masses; NGC 5194, a barred, \hm-rich spiral galaxy. The scale lengths $\rd$ and the densities $\Sigmastd$ of these galaxies were estimated by fitting exponential functions to $\Sigmaha(r)+\Sigmahm(r)$ for each galaxy (NGC 4214/5055/5194: $\fc=0.1/11/34$, $\rd=2.3/5.1/2.6\rm~kpc$, $\Sigmastd=10/56/139~\msunpc$). The good fit between the model and the observations demonstrates the validity of the pressure-based model for the \hm/\ha-ratio.

Eqs.~(\ref{eqsigmaha}, \ref{eqsigmahm}) make a list of predictions, which can be seen as direct consequences of the \hm/\ha--pressure relation: (i) the \hm/\ha-mass ratio of the whole galaxy increases with $\fc$ \citep{Obreschkow2009b}; (ii) \hm-rich galaxies, i.e.~$\fc\gtrsim10$, have annular \ha-distributions with a density drop towards the centre; (iii) if $\Sigmaha(r)$ and $\Sigmahm(r)$ cross, they do so close to the maximum of $\Sigmaha(r)$.

To evaluate $\Sigmaha(r)$ and $\Sigmahm(r)$ using Eqs.~(\ref{eqsigmaha}--\ref{eqfc}), we require an estimate of $\rd$, $\msdisk$, and $\mg$. In \citet{Obreschkow2009b}, we therefore adopted the cosmological galaxy simulation performed by \citet{DeLucia2007} on the dark-matter skeleton of the Millennium Simulation (\citealp{Springel2005}). In this ``semi-analytic'' simulation, galaxies were represented by a list of global properties, such as position and total masses of gas, stars, and black holes. These properties were evolved using simplistic formulae for mechanisms, such as gas accretion by infall and mergers, star formation, gas heating by supernovae, and feedback from black holes. The free parameters in this model were tuned mostly to optical observations in the local universe (see \citealp{Croton2006}), such as the joint luminosity/colour/morphology distribution of optically observed low-redshift galaxies. However, no measurements of \ha~and \hm~were used to adjust the free parameters.

The semi-analytic simulation resulted in a catalog listing the properties of $\sim3\cdot10^7$ model-galaxies at 64 cosmic time steps. In these galaxies, the cold gas was treated as a single component, hence masking the complexity of atomic and molecular phases. We therefore post-processed this simulation using Eqs.~(\ref{eqsigmaha}--\ref{eqfc}) to evaluate \ha- and \hm-distributions for every galaxy. The results of the emerging \ha~and \hm~simulations were presented in \citet{Obreschkow2009b}. They well match the \ha- and \hm-mass functions, mass--diameter relations, and mass--velocity relations observed in the local Universe. The high-$z$ predictions are roughly consistent with the sparse cold gas detections at $z>0$ \citep{Obreschkow2009c}.

The limitations and uncertainties of this model for \ha~and \hm~at low $z$ and high $z$ were discussed in detail in Section 6 of \citet{Obreschkow2009b}. Some crucial points shall be emphasized here: (i) The predicted space density of \ha,~$\Omegaha$, undershoots the inferences of damped Lyman-$\alpha$ systems (DLAs) at $z>1$ by about a factor 2 \citep[see Fig.~2 and Section 4 in][]{Obreschkow2009c}. We discussed possible reasons for this discrepancy, but it should be stressed that an offset by a factor 2 is rather small, given that the semi-analytic models of \cite{DeLucia2007} have never been tuned to \ha-data, but only to local optical data. (ii) One might worry whether the \hm/\ha--pressure relation, assumed constant in our model, could undergo a significant evolution with redshift. In Section 6.3 of \cite{Obreschkow2009b}, we argued that this effect is reasonably small because the empirical \hm/\ha--pressure relation seems to implicitly include changes in the photo-dissociating UV-radiation field, and the scatter in \hm/\ha~shows no significant correlation with metallicity \citep{Blitz2006}, although the mean metallicities of the galaxies in the sample differ by a factor 5. This factor is on the order of the typical metallicity evolution of an individual massive galaxy between redshift $z=10$ and $z=0$, and hence the cosmic evolution of metallicity seems to be a minor concern \citep[for details and references see Section 6.3 in][]{Obreschkow2009b}. (iii) Eq.~\ref{eqsigmaha} suggests that $\Sigmaha(r)$ falls of exponentially at large galactic radii, while some observations have suggested that it varies rather as $1/r$, proportional to the dark matter density derived from flat rotation curves \citep[e.g.][]{Bosma1981,Hoekstra2001}. The detailed \ha-maps and CO-maps presented by \cite{Leroy2008} for a sample of 18 nearby galaxies nevertheless reveal that all these galaxies carry most of their cold gas mass inside a radius where the angularly averaged \ha-distribution is consistent with an exponential profile. Therefore, we consider the exponential profile as a safe assumption for the purpose of splitting the cold gas mass between \ha~and \hm.

We use sample of 18 nearby galaxies, for which \citet{Leroy2008} derived inclination-corrected column density maps of \ha, \hm, stars, and SFRs (not used here).

\section{Cosmic evolution of $\Sigmaha$ and $\Sigmahm$}\label{section_global_evolution}

In this section, we shall investigate the average cosmic evolution of the surface densities of \ha~and \hm~in the simulated galaxies (see Section \ref{section_model}). We define the ``average'' $\avg{...}$ as the $(\mha+\mhm)$-weighted geometric average over all galaxies with $\mha+\mhm\geq10^8\msun$. The threshold $\mha+\mhm=10^8\msun$ approximately marks the completeness limit of the simulation and the mass-weighting ensures that massive galaxies with low space densities contribute significantly to the average.

\begin{figure}
  \includegraphics[width=\columnwidth]{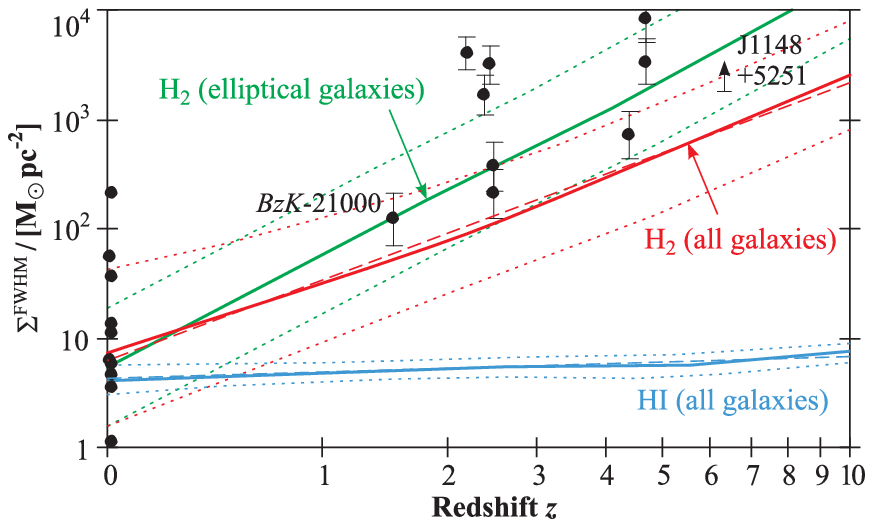}
  \caption{Solid and dotted lines show the simulated cosmic evolution of the surface densities $\avg{\Sigmahafwhm}$ and $\avg{\Sigmahmfwhm}$ with 1-$\sigma$ scatter. These lines represent interpolations between 51 discrete redshifts between $z=0$ and $z=10$. At each redshift, the average value relies on $\sim10^6-3\cdot10^7$ galaxies (the precise number depends on $z$). Dashed lines are the power law fits of Eqs.~(\ref{eqscaling3}, \ref{eqscaling4}). Points show the empirical data of Table \ref{tab_resolved_gals}.}
  \label{fig_global_evolution}
\end{figure}

For each galaxy we define the Full-Width-Half-Maximum (FWHM) surface densities, $\Sigmahafwhm$ and $\Sigmahmfwhm$, as the average surface densities inside the radii $r$, where $\Sigmaha(r)$ and $\Sigmahm(r)$ reach $50\%$ of their maximal value. The simulated cosmic evolution of $\avg{\Sigmahafwhm}$ and $\avg{\Sigmahmfwhm}$ is displayed in Fig.~\ref{fig_global_evolution} as solid lines. Both $\avg{\Sigmahafwhm}(z)$ and $\avg{\Sigmahmfwhm}(z)$ are well fitted by power-laws (dashed lines in Fig.~\ref{fig_global_evolution}),
\bea
  \avg{\Sigmahafwhm/[\msunpc]}(z) &~=~& 4.7\,(1+z)^{0.2},  \label{eqscaling3} \\
  \avg{\Sigmahmfwhm/[\msunpc]}(z) &~=~& 7.1\,(1+z)^{2.4}.  \label{eqscaling4}
\eea

The simulated \ha-density remains roughly constant reaching maximal values around $8-9~\msunpc$, consistent with the observed \ha-saturation level in nearby galaxies \citep{Bigiel2008}. This cosmic evolution of $\Sigmahmfwhm$ is a pure prediction, with no observational counterpart at $z>0$ to-date.

The simulated \hm-densities can be compared to CO-data. The data points in Fig.~\ref{fig_global_evolution} (see also Table \ref{tab_resolved_gals}) represent 12 local and 11 distant galaxies with observational estimates of $\Sigmahmfwhm$ derived from resolved CO-maps\footnote{The adopted CO-to-\hm~conversion is $\xco=4{\rm~K\,km\,s^{-1}\,pc^2}$ at $z=0$ (Table \ref{tab_resolved_gals}, top), consistent with observations in local quiescent galaxies \citep{Leroy2008}; and $\xco=1{\rm~K\,km\,s^{-1}\,pc^2}$ for sources at $z>0$ (Table \ref{tab_resolved_gals}, bottom), consistent with observations in Ultra Luminous Infrared Galaxies \citep{Downes1998}.}. To our knowledge, none of the high-redshift galaxies are gravitationally lensed, however they certainly represent a highly-biased sample (see below).

At $z=0$, the observed average of $\Sigmahmfwhm$, weighted by the cold gas masses and the space densities drawn from the cold gas mass function \citep{Obreschkow2009a}, is $\avg{\Sigmahmfwhm}={7\pm1}~\msun\,{\rm pc}^{-2}$, consistent with the simulated value of $\avg{\Sigmahmfwhm}={7.9}~\msun\,{\rm pc}^{-2}$.

At $z\geqslant0$, the empirical values of $\avg{\Sigmahmfwhm}$ show a clear trend to increase with redshift $z$. We must worry whether this is a pure selection effect, merely reflecting the relation between sensitivity and redshift \citep[similar to Fig.~5 in][]{Greve2005}. To our knowledge, all the high-redshift CO-sources considered here have originally been selected from single-mirror surveys (submillimeter, IR, optical), which are flux limited rather than surface brightness limited. Unfortunately, it is not clear what fraction of all flux selected sources has been detected in the follow-up interferometric surveys, but it is likely that the present sample of spatially resolved high-redshift CO-sources is biased towards high fluxes (i.e.~large \hm-masses) rather than towards high surface brightness (i.e.~high \hm-surface densities). In general, more massive disc galaxies tend to have lower surface densities than less massive ones. Therefore, the selection bias of a flux limited sample tends to bias high-redshift disc galaxies towards lower \hm-surface densities. If the observed systems were single discs, one would thus expect the real relation between \hm-surface densities and redshift to be even steeper than suggested by the data points in Fig.~\ref{fig_global_evolution}. On the other hand, the CO-detected galaxies at $z>0$ seem to be heavily biased towards systems subjected to major mergers \citep{Tacconi2006}. On average, major mergers decrease the specific angular momentum, hence decreasing $\rd$ and increasing $\Sigmahmfwhm$.

\begin{table}
\centering
\begin{tabular*}{\columnwidth}{p{3.2cm}p{0.9cm}p{0.4cm}p{0.8cm}p{1.7cm}}
\hline
{\bf Object} & {\bf Type} & {$\mathbf{z}$} & {\bf CO-}  & {$\mathbf{\log(\Sigmahmfwhm/}$} \\
             &            &                & {\bf line} & {$\mathbf{[\msunpc])}$} \\
\hline
   NGC0628 &         Sc &        0.0 &        2--1 &  $\!\!\!\!-0.4\pm0.1^{\rm(a)}$ \\
   NGC3198 &        SBc &        0.0 &        2--1 &  $0.7\pm0.1^{\rm(a)}$ \\
   NGC3184 &        SBc &        0.0 &        2--1 &  $0.6\pm0.1^{\rm(a)}$ \\
   NGC4736 &        Sab &        0.0 &        2--1 &  $1.2\pm0.1^{\rm(a)}$ \\
   NGC3351 &        SBb &        0.0 &        2--1 &  $0.1\pm0.1^{\rm(a)}$ \\
   NGC6946 &        SBc &        0.0 &        2--1 &  $2.4\pm0.1^{\rm(a)}$ \\
   NGC3627 &        SBb &        0.0 &        2--1 &  $1.6\pm0.1^{\rm(a)}$ \\
   NGC5194 &        SBc &        0.0 &        2--1 &  $1.1\pm0.1^{\rm(a)}$ \\
   NGC3521 &       SBbc &        0.0 &        2--1 &  $0.8\pm0.1^{\rm(a)}$ \\
   NGC2841 &         Sb &        0.0 &        2--1 &  $\!\!\!\!-0.1\pm0.1^{\rm(a)}$ \\
   NGC5055 &        Sbc &        0.0 &        2--1 &  $1.8\pm0.1^{\rm(a)}$ \\
   NGC7331 &        SAb &        0.0 &        2--1 &  $0.8\pm0.1^{\rm(a)}$ \\
\hline
 \textit{BzK}-21000  &  Galaxy &        1.5 &        2--1 &  $2.1\pm0.2^{\rm(b)}$ \\
 SMM J123549+6215    &  SMG    &        2.2 &        3--2 &  $3.6\pm0.1^{\rm(c)}$ \\
 SMM J163650+4057    &  SMG    &        2.4 &        3--2 &  $3.3\pm0.2^{\rm(c)}$ \\
 SMM J163658+4105    &  SMG    &        2.5 &        3--2 &  $3.5\pm0.2^{\rm(c)}$ \\
 SMM J123707+6214SW  &  SMG    &        2.5 &        3--2 &  $2.4\pm0.2^{\rm(c)}$ \\
 SMM J123707+6214NE  &  SMG    &        2.5 &        3--2 &  $2.6\pm0.2^{\rm(c)}$ \\
 BRI 1335 0417       &  QSO    &        4.4 &        2--1 &  $2.9\pm0.2^{\rm(d)}$ \\
 BRI 1202-0725 north &  QSO    &        4.7 &        2--1 &  $3.6\pm0.2^{\rm(d)}$ \\
 BRI 1202-0725 south &  QSO    &        4.7 &        2--1 &  $4.0\pm0.2^{\rm(d)}$ \\
 J1148+5251          &  QSO    &        6.4 &        3--2 &  $>3.6^{\rm(e)}$ \\
\hline
\end{tabular*}
   \caption{Local (top) and distant (bottom) galaxies with spatially resolved CO-detections, and corresponding estimated \hm-surface densities $\Sigmahmfwhm$. (a) From radial \hm-density profiles given in \citet{Leroy2008}; (b) from CO-fluxes and velocity peak separation measured by \citet{Daddi2008}; (c) from Table 1 in \citet{Tacconi2006}, for the submillimeter galaxy (SMG) SMM J163650+4057 the radius was averaged between semi-major and semi-minor axis, for SMM J123549+6215 the CO-flux was determined from CO(3--2) and the radius from CO(6--5); (d) from \citet{Carilli2002}, BRI 1335 0417 probably has two non-resolved sub-components; (e) \citet{Walter2004}.}
   \label{tab_resolved_gals}
\end{table}

We should therefore compare the CO-measurements at $z>0$ exclusively with the simulated galaxies with major mergers in their cosmic evolution. In the semi-analytic model, the evolution of the angular momentum during a merger is directly computed from the underlying $N$-body Millennium Simulation. We can therefore expect that, on average, mergers decrease the specific angular momentum, and hence increase the cold gas densities towards the high observed densities shown in Fig.~\ref{fig_global_evolution}. Within the semi-analytic model \citep{Croton2006}, where stellar bulges arise during mergers, galaxies resulting from major mergers are ``elliptical'' galaxies, defined as the objects with a bulge-to-total mass ratio larger than $0.4$ (see eq.~18 in \citealp{Obreschkow2009b}). The average $\avg{\Sigmahmfwhm}$ and the corresponding 1-$\sigma$ scatter of the simulated elliptical galaxies are represented by the green lines in Fig.~\ref{fig_global_evolution} and provide a much better fit to the observational data.

Despite the reasonable agreement between the simulated and the empirical data for $\avg{\Sigmahafwhm}$ and $\avg{\Sigmahmfwhm}$, we emphasize that there are many subtle selection effects in the empirical data (e.g.~flux limited sample versus surface brightness limited sample, Eddington bias) and that the semi-analytic models remain in many regards simplistic and uncertain (e.g.~no geometrical model of mergers, mass resolution limitations, no ram-pressure stripping). Clearly, the advent of telescopes, such as the SKA, the LMT, and ALMA, will prove invaluable to refine these models.

What are the reasons for the different cosmic evolutions of $\Sigmaha$ and $\Sigmahm$? According to Eqs.~(\ref{eqsigmaha}--\ref{eqfc}), these evolutions can be understood from the average cosmic evolution of $\rd$, $\mg$, and $\ms$. Most of the massive galaxies in the semi-analytic simulation are gas-dominated at $z>1$ and their cold gas masses remain roughly constant with cosmic time due to a self-regulated equilibrium between the net cold gas accretion and star formation. Thus, for any given galaxy, the cosmic evolution of the value $\fc$ and the functions $\Sigmaha(r)$ and $\Sigmahm(r)$ is essentially dictated by the evolution of the scale radius $\rd$. Assuming a similar specific angular momentum for the galaxy and its halo \citep{Fall1980}, $\rd$ is expected to evolve proportionally to the virial radius $\rvir$ of the halo. For a spherical halo of mass $\mvir$, the latter scales as $\rvir^3\propto\mvir/[\Omega_{\rm m}(1+z)^3+\Omega_\Lambda\big]$ \citep[flat universe,][]{Gunn1972}, and hence for a fixed $\mvir$,
\be\label{eqrdevol}
  \rd\propto[\Omega_{\rm m}(1+z)^3+\Omega_\Lambda\big]^{-1/3},
\ee
where $\Omega_{\rm m}$ (here 0.25) and $\Omega_\Lambda$ (here 0.75) denote the normalized space densities of matter and vacuum energy.

At high $z$, Eq.~(\ref{eqrdevol}) reduces to $\rd\propto(1+z)^{-1}$, consistent with observations in the Hubble Ultra Deep Field \citep{Bouwens2004}. If the cosmic evolution of $\mg$ and $\ms$ is neglected, then $\rd\propto(1+z)^{-1}$ implies $\fc\propto(1+z)^{3.2}$ (see Eq.~\ref{eqfc}) and $\Sigmastd\propto(1+z)^2$. Yet, for $\fc>5/3$, the maximum of $\Sigmaha(r)$ is given by $\Sigmahamax=0.516\,\Sigmastd\,{\fc}^{-5/8}$ \citep[see][]{Obreschkow2009b} and hence $\Sigmahamax\propto(1+z)^2\,(1+z)^{-2}=\rm const$. In other words, the surface density of \ha~is expected to show little evolution with redshift, consistent with the numerical fit for $\avg{\Sigmahafwhm}(z)$ of Eq.~(\ref{eqscaling3}). On the other hand, $\Sigmahm(r)$ in Eq.~(\ref{eqsigmahm}) reduces to $\Sigmastd\,\exp(-r/\rd)$ $\forall\,r>0$, if $\fc\gg1$ (i.e.~$z\gg1$). Hence, if the cosmic evolution of $\mg$ is negligible, $\Sigmahmmax$ and $\Sigmahmfwhm$ are predicted to scale as $\propto\Sigmastd\propto(1+z)^2$. A comparison to Eq.~(\ref{eqscaling4}) confirms that, within our model, the cosmic evolution of $\Sigmahm$ is largely explained by the size-evolution of galaxies, while the evolution of $\mg$ plays a minor role, accounting for an additional $\propto(1+z)^{0.4}$ scaling.

\section{Evolution Scenario for the Milky Way}\label{section_mw_evolution}

We shall now investigate the cosmic evolution of the \ha-~and \hm-distributions in MW-type galaxies. By definition, a simulated galaxy at $z=0$ is called a ``MW-type'' galaxy, if its morphological type is Sb--Sc and if it matches the stellar mass $\ms$, the \ha-mass $\mha$, the \hm-mass $\mhm$, the \ha-half-mass radius $\rhahalfmass$, and the \hm-half-mass radius $\rhmhalfmass$ of the MW within a factor 1.3. This factor roughly matches the empirical uncertainties of the MW data in Table \ref{tab_case_study}. Within this definition, the simulation contains $2\cdot10^3$ MW-type galaxies at $z=0$. At $z>0$, we define MW-type galaxies as those objects, which are the most massive progenitors of a MW-type galaxy at $z=0$. Most MW-type galaxies accrete about half of their mass in a time-interval corresponding to the redshift range $z=2-10$. They typically undergo a series of minor mergers, allowing the build-up of small bulges.

To accommodate the nature of the empirical data (e.g.~poor data for gas at the MW centre), we here consider the half-mass radii $\rhahalfmass$ and $\rhmhalfmass$, and the enclosed average surface densities $\Sigmahahalfmass$ and $\Sigmahmhalfmass$, rather than FWHM values.

\begin{figure}
  \includegraphics[width=\columnwidth]{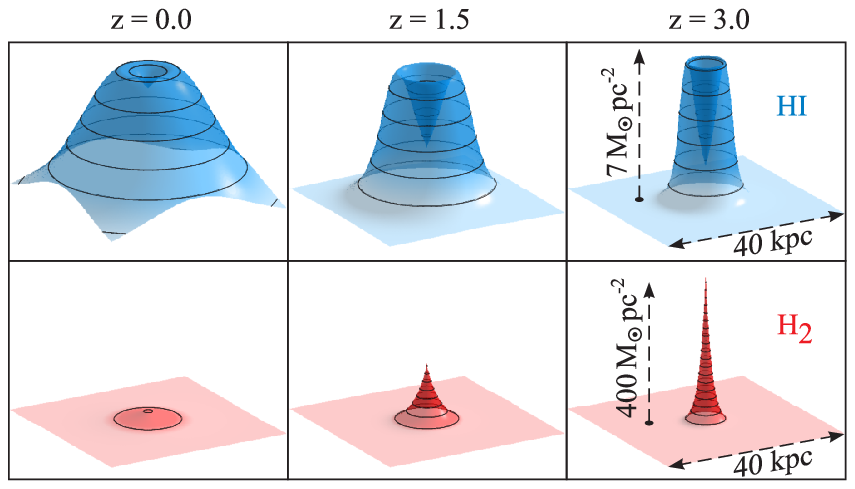}
  \caption{Cosmic evolution of the average profiles $\Sigmaha(r)$ and $\Sigmahm(r)$ calculated from $2\cdot10^3$ simulated MW-type galaxies. The contours are spaced by $1\,\msun{\rm\,pc}^{-2}$ for \ha~and $20\,\msun{\rm\,pc}^{-2}$ for \hm, with the bottom contour corresponding to the separation value. The redshifts $z=1.5$ and $z=3$ respectively correspond to the galaxy $BzK$-21000 \citep{Daddi2008} and to the primary science goal of ALMA, i.e.~the detection of MW-type galaxies at $z=3$ in less than 24 hours observation.}
  \label{fig_case_study}
\end{figure}

In Table \ref{tab_case_study}, we explicitly present the predicted average gas masses, gas radii, and gas densities of the $2\cdot10^3$ MW-type galaxies at three specific redshifts: $z=0$, corresponding to the MW itself; $z=1.5$, corresponding to the galaxy $BzK$-21000, which is the only ``relatively ordinary'' galaxy (i.e.~a borderline ultra luminous infrared galaxy for a ``normal'' star formation efficiency) at higher $z$ detected and spatially resolved in CO-emission to-date; $z=3$, corresponding to the primary science goal of ALMA, i.e.~the detection of MW-type galaxies at $z=3$ in less than 24 hours observation. Fig.~\ref{fig_case_study} illustrates the average density profiles $\Sigmaha(r)$ and $\Sigmahm(r)$ at these three redshifts.

The cosmic evolution of the average \ha- and \hm-properties of the simulated MW-type galaxies can be characterized as follows: (i) the mass ratio $\mhm/\mha$ approximately varies as $(1+z)^{1.6}$, identical to the space density evolution $\Omegahm/\Omegaha\propto(1+z)^{1.6}$ predicted in \citet{Obreschkow2009c}; (ii) the surface density of \ha~remains approximately constant, while the density of \hm~increases by a factor 30 from $z=0$ to $z=3$; (iii) both the \ha-radius and the \hm-radius shrink by a factor $4-5$ from $z=0$ to $z=3$; (iv) the \ha-distribution is more annular at high $z$ (see Fig.~\ref{fig_case_study}), resembling the central \ha-deficiency seen in \hm-rich local galaxies.

\begin{table}
  \centering
\small{\begin{tabular*}{\columnwidth}{p{2.1cm}p{2.1cm}p{2.0cm}p{1.2cm}}
\hline & Sim/MW & Sim/$BzK$ & Sim \\ \hline
Redshift $z$ & \mbox{$0.0$} & \mbox{$1.5$} & \mbox{$3.0$} \\
Type & \mbox{Sb-c/SBbc} & \mbox{Sc/S?} & \mbox{Sc-d} \\
$\ms/[10^{10}\msun]$ & \mbox{$5^{+0.3}_{-0.2}$/$5^{+1}_{-1}{^{\rm(a)}}$} & \mbox{$1.7^{+0.2}_{-0.2}$/$6^{+1}_{-1}$} & \mbox{$0.5^{+0.1}_{-0.1}$} \\
$\mha/[10^{9}\msun]$ & \mbox{$7.8^{+0.4}_{-0.3}$/$8^{+2}_{-2}{^{\rm(b)}}$} & \mbox{$2.3^{+0.9}_{-0.5}$/?} & \mbox{$0.7^{+0.3}_{-0.2}$} \\
$\rhahalfmass/[\kpc]$ & \mbox{$16^{+0.5}_{-0.4}$/$15^{+5}_{-5}{^{\rm(b)}}$} & \mbox{$7.3^{+1.2}_{-0.8}$/?} & \mbox{$3.9^{+0.7}_{-0.5}$} \\
$\Sigmahahalfmass/[\msun{\rm pc^{-2}}]$ & \mbox{$4.5^{+0.1}_{-0.1}$/$6^{+3}_{-3}{^{\rm(b)}}$} & \mbox{$5.4^{+0.3}_{-0.3}$/?} & \mbox{$5.5^{+0.4}_{-0.4}$} \\
$\mhm/[10^{9}\msun]$ & \mbox{$3.5^{+0.2}_{-0.2}$/$3.5^{+1}_{-1}{^{\rm(c)}}$} & \mbox{$4.6^{+0.7}_{-0.5}$/$17^{+5}_{-5}$} & \mbox{$2.9^{+0.7}_{-0.4}$} \\
$\rhmhalfmass/[\kpc]$ & \mbox{$6.8^{+0.2}_{-0.2}$/$7^{+1}_{-1}{^{\rm(c)}}$} & \mbox{$2.9^{+0.5}_{-0.4}$/$4^{+1}_{-1}$} & \mbox{$1.4^{+0.3}_{-0.2}$} \\
$\Sigmahmhalfmass/[\msun{\rm pc^{-2}}]$ & \mbox{$12^{+1}_{-1}$/$11^{+3}_{-3}{^{\rm(c)}}$} & \mbox{$130^{+50}_{-20}$/$150^{+50}_{-50}$} & \mbox{$360^{+190}_{-70}$} \\
\hline
\end{tabular*}}
   \caption{Average value and 1-$\sigma$ scatter of the cold gas and stellar properties of $2\cdot10^3$ simulated MW-type galaxies (Sim) at $z=0$, $z=1.5$, and $z=3$. The actual measurements of the MW with 1-$\sigma$ uncertainties were adopted from: (a) \citet{Flynn2006}; (b) from analytic fits to $\Sigmaha(r)$ in \citet{Kalberla2008}; (c) from $\Sigmahm(r)$ in Table 3 in \citet{Sanders1984}. The properties of the simulated MW-type galaxies at $z=1.5$ are compared to those of the CO-detected galaxy $BzK$-21000 \citep{Daddi2008}.}
   \label{tab_case_study}
\end{table}

The comparison of the galaxy $BzK$-21000 \citep{Daddi2008} and the MW-type galaxies at $z=1.5$ (see Table \ref{tab_case_study}) suggests that the \hm-density of $BzK$-21000 is consistent with that of a typical MW progenitor, while the total \hm-mass of $BzK$-21000 may be a few times higher. Considering that the MW lies at the lower end of intermediate mass spiral galaxies (stellar masses $\ms$ in the range $3\cdot10^{10}-3\cdot10^{11}~\msun$; \citealp{Flynn2006}), the \hm-distribution of $BzK$-21000 appears to be typical for the progenitors of intermediate mass disc galaxies.

\section{Conclusion}\label{section_conclusion}

We have studied the cosmic evolution of the surface densities of \ha~and \hm~in regular galaxies using theoretical models combined with the Millennium Simulation.

A key result is that the surface density of \ha~remains approximately constant and close to saturation at all redshifts, while the mean surface density of \hm~changes dramatically as $\propto(1+z)^{2.4}$, mainly due to the size evolution of galaxies. These predictions will become testable with future telescopes, such as the LMT, ALMA, and the SKA. The few CO-detected high-$z$ galaxies available today seem to have even higher \hm-surface densities than predicted by the $\propto(1+z)^{2.4}$ scaling (see Fig.~\ref{fig_global_evolution}). This could result from a selection bias towards systems, which have lost some of their specific angular momentum in major mergers.

We also studied the cosmic evolution of the cold gas in a sample of MW-type galaxies. We predicted that at $z=3$ the \hm-mass of a MW-type galaxy is $\mhm\approx3\cdot10^9\,\msun$ with the denser 50\% of this mass reaching an average surface density of $\Sigmahmhalfmass\approx300-500\,\msun{\rm pc^{-2}}$.

\vspace{0.6cm}
We thank the anonymous referee for a list of very helpful suggestions. This activity is supported by the European Community Framework Programme 6, Square Kilometre Array Design Studies, contract no 011938. The Millennium Simulation databases were built as part of the activities of the German Astrophysical Virtual Observatory.


\label{lastpage}

\end{document}